\documentclass{appolb}
\usepackage{graphicx}
\usepackage{lineno}
\usepackage{subcaption}
\usepackage{float}
\usepackage{tikz}
\usepackage{placeins}
\usepackage{amsmath,amssymb}

\begin{document}
\title{Measurements of particle spectra in diffractive
proton-proton collisions with the STAR detector at RHIC\thanks{Presented at Diffraction and Low-$x$ 2018 workshop, 26 August 2018 to 1 September 2018, Reggio Calabria, Italy}%
}
\author{{\L}ukasz Fulek\thanks{This work was partially supported by the Polish National Science Centre grant No. UMO-2015/18/M/ST2/00162 and by the AGH UST grant No. 15.11.220.717/28 for 
young researchers within subsidy of Ministry of Science and Higher Education.} for the STAR Collaboration
\address{AGH University of Science and Technology,\\Faculty of Physics and Applied Computer Science,\\ al. A. Mickiewicza 30, 30-059 Krak\'ow, Poland
}
}
\maketitle
\begin{abstract}
\noindent
We present the inclusive and identified particle (pion, proton and their antiparticle) production in Single Diffraction Dissociation and Central Diffraction processes with the STAR detector at RHIC. 
The~forward-scattered proton(s) are tagged in the STAR Roman Pot system while the~charged particle tracks are reconstructed in the~STAR Time Projection Chamber. 
Ionization energy loss  and time of flight of charged particles are used for particle identification. In addition, the~proton-antiproton production asymmetry as a~function of rapidity is measured and used to  study the~baryon number transfer over a~large rapidity interval in Single Diffraction Dissociation. 
\end{abstract}
  
\section{Introduction}
Diffractive processes at high energies are characterized by the exchange of a color singlet object with vacuum quantum numbers, so called Pomeron, and are well described by the Regge theory.  
In $p$+$p$ collisions two kind of processes are of particular interest,   Single Diffraction Dissociation (SD: $p+p\to p+X$) and Central  Diffraction $\left(\textnormal{CD: }p+p\to p+X+p\right)$,  where $X$ denotes the diffractively produced system. 

Charged-particle measurements in $p$+$p$ collisions provide insight into the strong interaction in the low-energy, non-perturbative regime of QCD. Particle interactions at these energy scales are typically described by QCD-inspired models implemented in Monte Carlo (MC) event generators with free parameters that can be constrained by  measurements. Charged-particle distributions have been measured previously in minimum bias (MB) inelastic hadron-hadron collisions at
various center-of-mass energies~\cite{cms,atlaschrg}.
The identified charged particle production in the midrapidity region has been also widely studied starting from the very first experiments performed at ISR at CERN to contemporary measurements with very high center-of-mass energies at RHIC \cite{PhysRevC.79.034909} and LHC \cite{Adam:1999164}. 

In the Standard Model, the baryon 
number is a conserved quantity in all interactions. The conserved baryon number associated with the beam particles is called ,,baryon number transfer" and has been studied both theoretically~\cite{Kopeliovich:1988qm,Rossi:1977cy} and  experimentally in MB interactions~\cite{Aamodt:2010dx}, by a~long time. The similar effect can be also  studied in SD interactions, where the~direction of the initial baryon is uniquely defined. The baryon number transfer is quantified by the baryon to antibaryon ratios.

In this paper we report on measurements of inclusive charged-particle distributions and identified particle/antiparticle ratios as a function of transverse momentum $p_T$ and pseudorapidity $\eta$ in CD and SD $p$+$p$ collisions at $\sqrt{s}=200$~GeV using data collected by the STAR experiment in 2015. In addition, we focus on the~asymmetry of the production of protons and antiprotons at midrapidity in SD process. 

\section{Experimental setup}
The main components of the~STAR detector \cite{Ackermann:2002ad} used in this analysis include  the Time Projection Chamber~(TPC), which provides information about momentum  and ionization energy losses $\left(dE/dx\right)$ of charged particles, the Time-Of-Flight (TOF) system which extends the capability of TPC in particle identification and the Roman Pot detectors (RP) \cite{wlodek}, which enable tagging of the forward scattered protons. 
\section{Charged-particle distributions}
\begin{figure}[h]
	\hspace*{-1em}
	\parbox{0.47\textwidth}{
		\centering
		\begin{subfigure}[b]{\linewidth}{
				{\includegraphics[width=1.1\linewidth, page=1]{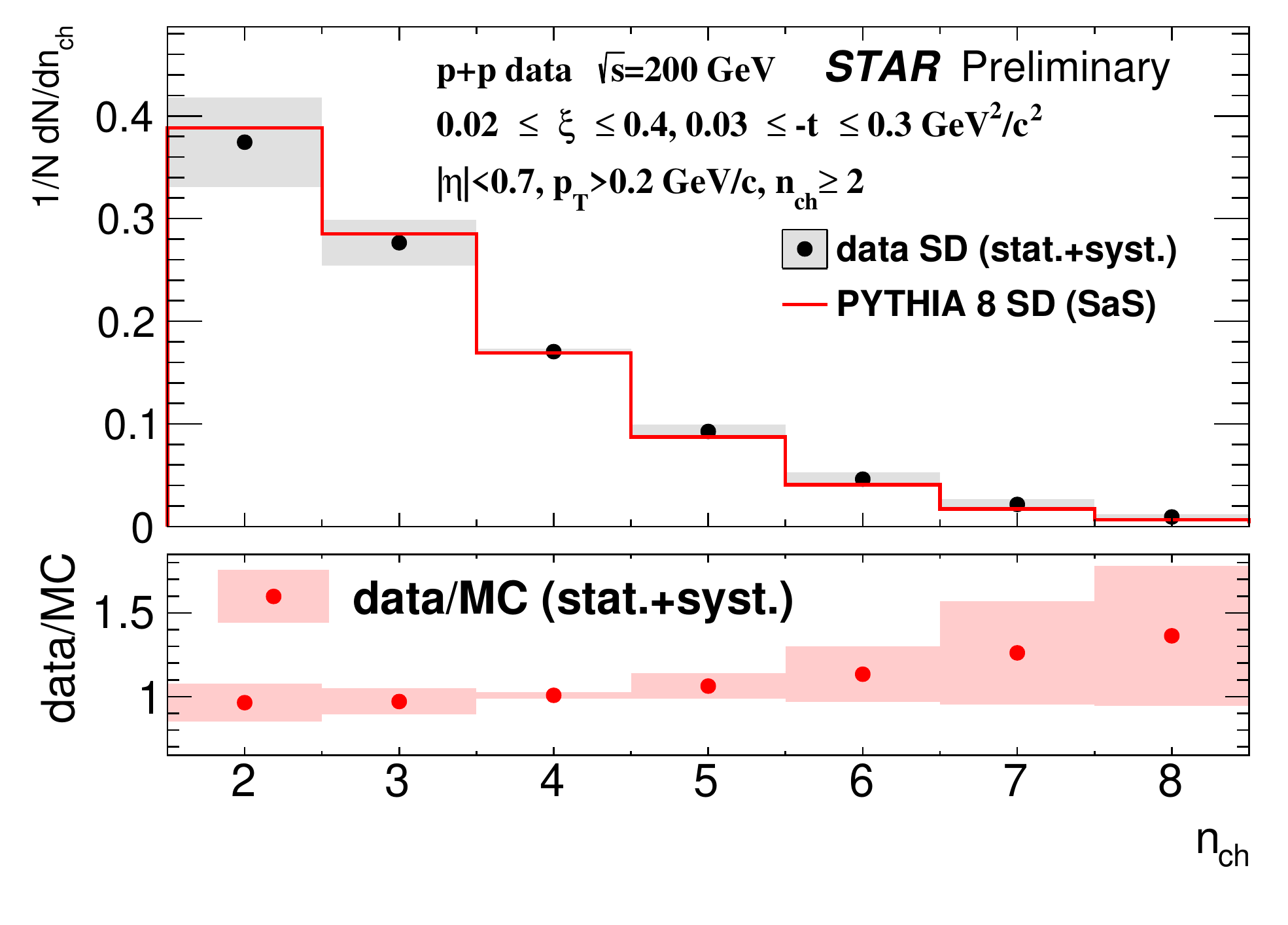}}}
		\end{subfigure}
	}
	\quad
	\parbox{0.47\textwidth}{
		\centering
		\begin{subfigure}[b]{\linewidth}{
				{\includegraphics[width=1.1\linewidth, page=1]{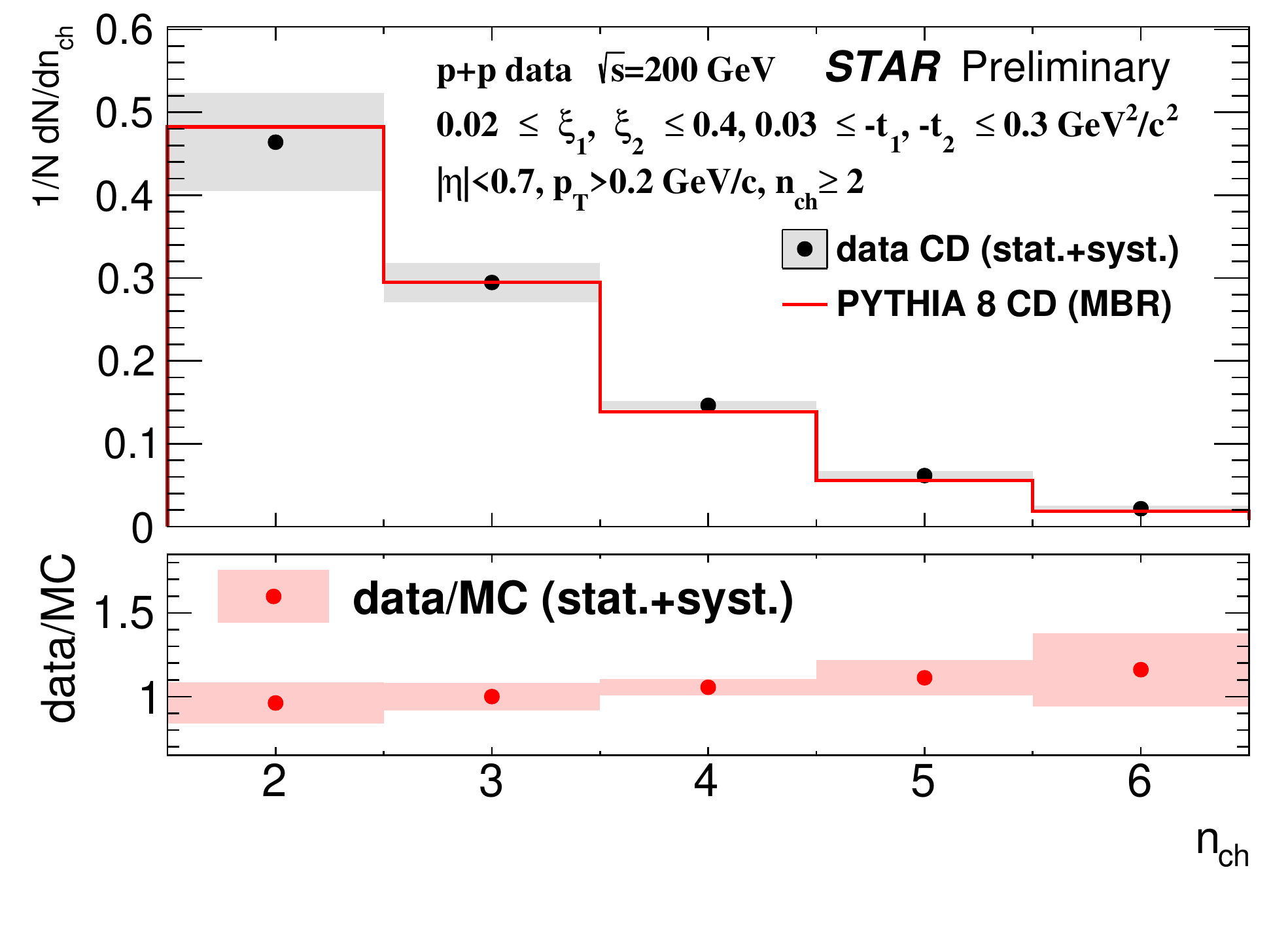}}}
		\end{subfigure}
	}
	\caption[]{Multiplicity distributions of primary charged particles for SD~(left) and CD~(right) processes. Data are compared to PYTHIA~8 Monte Carlo (MC) simulations.}
	\label{fig:nch}
\end{figure}

\begin{figure}
	\hspace*{-1em}
	\parbox{0.47\textwidth}{
		\centering
		\begin{subfigure}[b]{\linewidth}{
				{\includegraphics[width=1.1\linewidth, page=1]{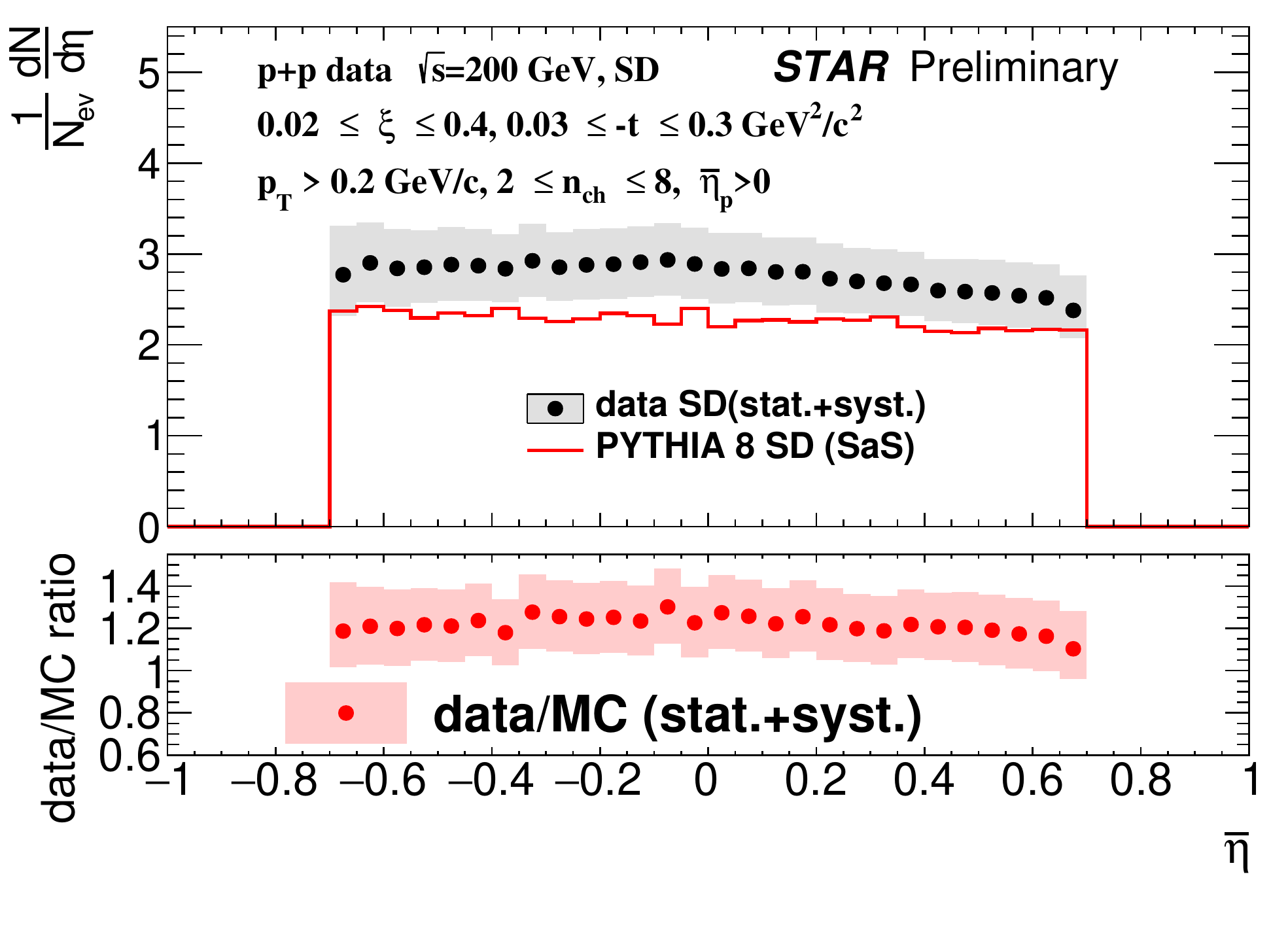}}}
		\end{subfigure}
	}
	\quad
	\parbox{0.47\textwidth}{
		\centering
		\begin{subfigure}[b]{\linewidth}{
				{\includegraphics[width=1.1\linewidth, page=1]{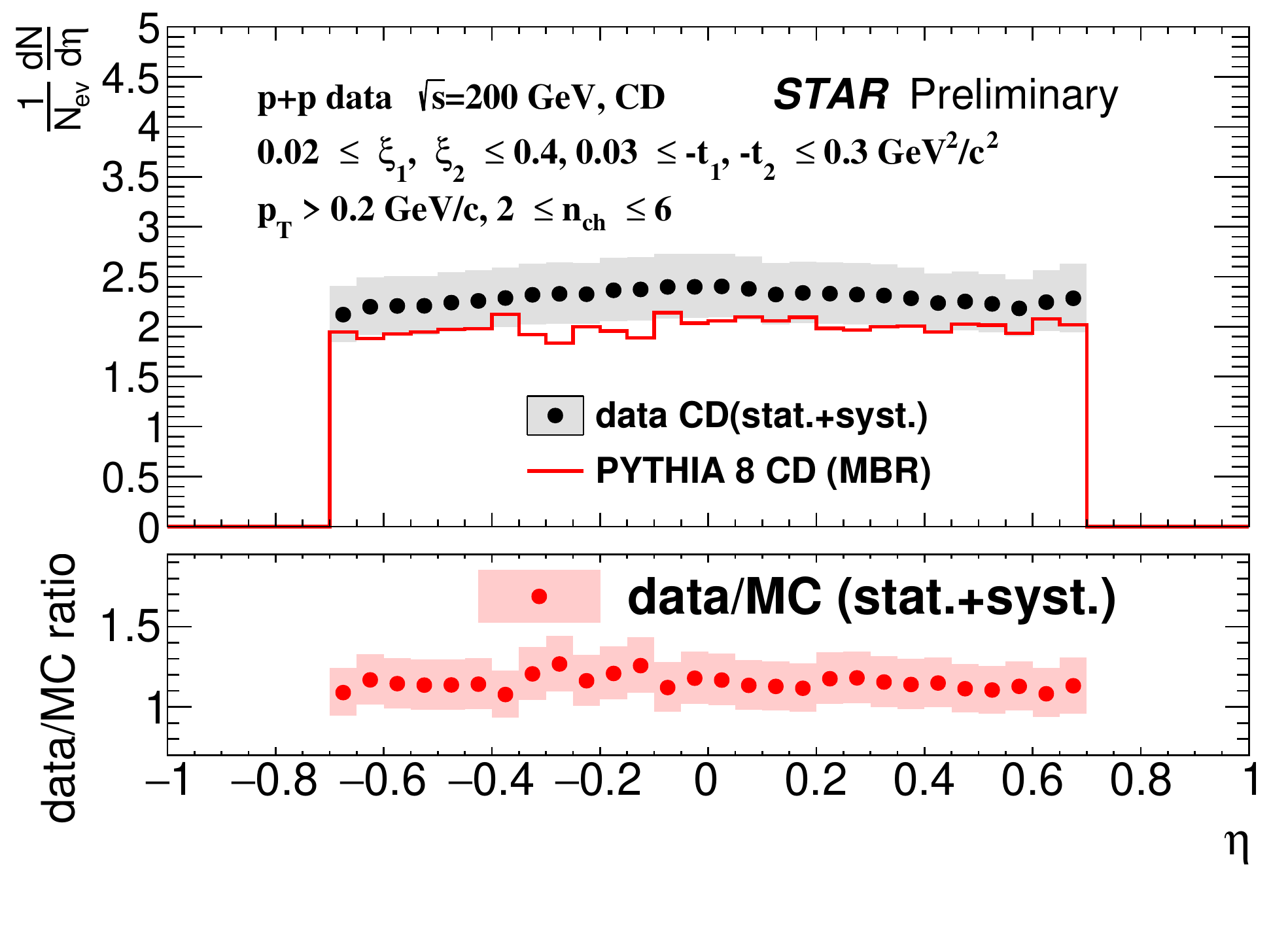}}}
		\end{subfigure}
	}
	\caption[]{Pseudorapidity ($\bar{\eta}$ for SD, $\eta$ for CD) dependence of the charged particles for SD~(left) and CD~(right). Data are compared to PYTHIA~8 MC simulations.}
	\label{fig:eta}
\end{figure}

The selection of diffractive events starts with the reconstruction of forward proton tracks. Exactly one track is 
required on each side of the Interaction Point (IP) for CD and exactly one track on one side only for SD (with fractional momentum loss $0.02 \leq\xi\leq0.4$). Only one primary vertex with $z$-component $|V_z|<80$~cm and at least two charged primary tracks with hits in the TOF system are required. To obtain the charged-particle multiplicity distributions, the distributions of tracks are unfolded into the distributions of particles using Bayes theorem. Figure~\ref{fig:nch} shows the measured distributions of primary charged particles  in the kinematic range $p_T > 200$~MeV/c and $|\eta| < 0.7$ for SD and CD processes together with PYTHIA 8.182~\cite{pythia8} prediction using SaS \cite{Schuler:1993wr} and MBR \cite{Ciesielski} Pomeron fluxes, respectively.
\begin{figure}[h]
	\hspace*{-1em}
	\parbox{0.47\textwidth}{
		
		\begin{subfigure}[b]{\linewidth}{
				{\includegraphics[width=1.1\linewidth, page=1]{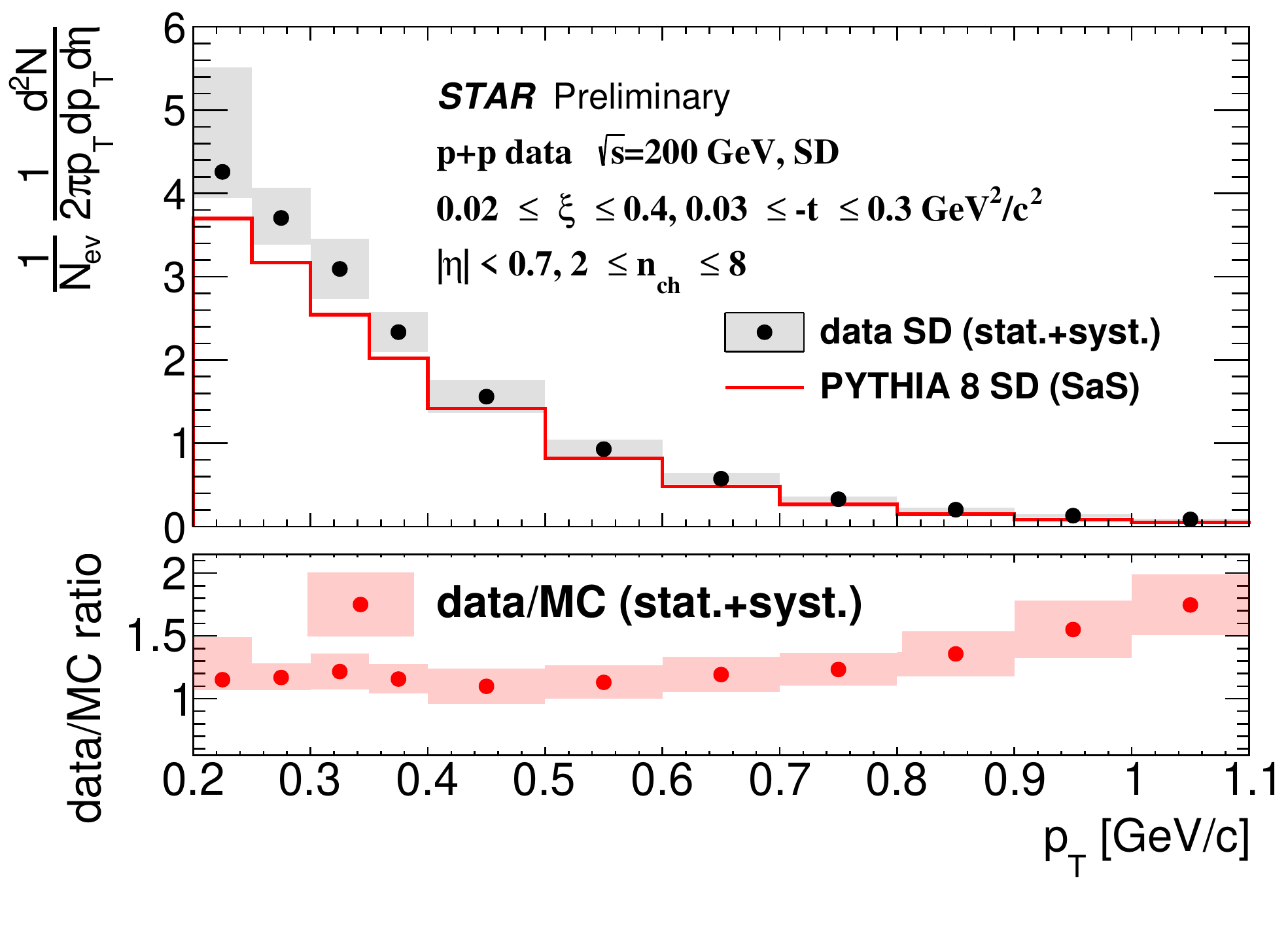}}}
		\end{subfigure}
	}
	\quad
	\parbox{0.47\textwidth}{
		
		\begin{subfigure}[b]{\linewidth}{
				{\includegraphics[width=1.1\linewidth, page=1]{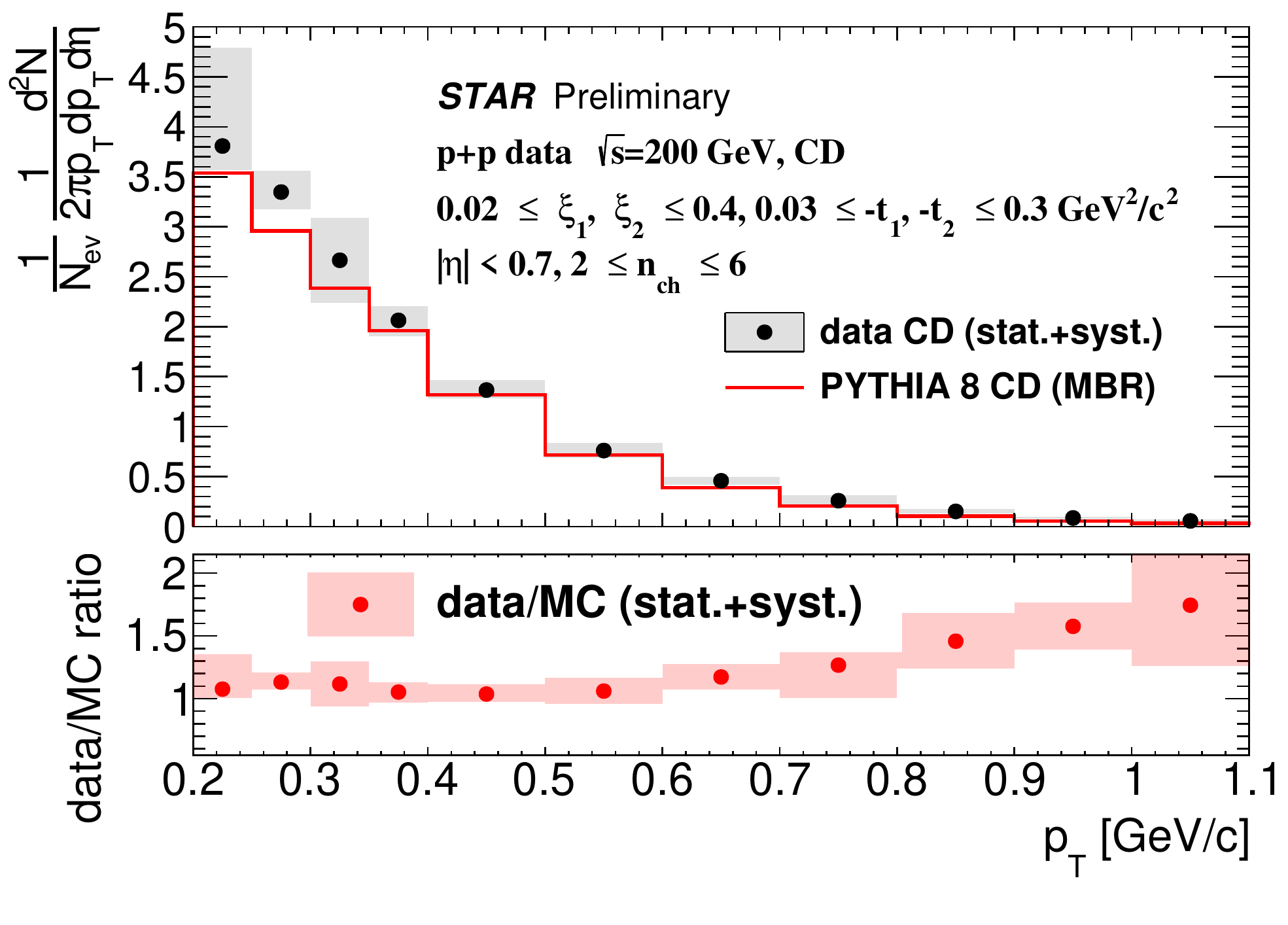}}}
		\end{subfigure}
	}
	\caption[]{ The rate of charged particles as a function of transverse momentum for SD~(left) and CD~(right) processes. Data are compared to PYTHIA~8 MC simulations.}
	\label{fig:pt}
\end{figure}
The analysis is limited to the number of charged particles~$n_{ch}$ between 2 and 8 for SD and $2 \leq n_{ch} \leq 6$ for CD, to avoid large background. The~measurements could be described by the~PYTHIA~8 simulation very well. Figure~\ref{fig:pt} shows the charged-particle transverse momentum distributions in SD and CD $p+p$ collisions. Generally, data are underestimated by PYTHIA 8  especially for relatively high-$p_T$ ($p_T>0.8$~GeV/c). Analysing the $\eta$ distribution require to define a new variable $\bar{\eta}=\frac{p_z^{proton}}{|p_z^{proton}|}\cdot\eta$ for SD, where $p_z^{proton}$ is the longitudinal momentum of the forward proton: positive $\bar{\eta}$ for particles produced on the outgoing proton direction and negative on the opposite side of the IP. Figure~\ref{fig:eta} shows the density of charged particles as a function of pseudorapidity. PYTHIA 8 simulations  describes the data within $5\%$ accuracy.

\section{Identified particle spectra}
In the case of the identified particle spectra measurements, we require  $|V_z|<100$~cm and $\xi<0.6$. The charged tracks are measured in the kinematic range $p_T > 200$~MeV/c and $|\eta| < 1$. The combined information from TPC and TOF is used to identify charged pions $\left(\pi^\pm\right)$, protons and antiprotons. 

\begin{figure}
	\hspace*{-1em}
	\parbox{0.5\textwidth}{
		\centering
		\begin{subfigure}[b]{\linewidth}{
				{\includegraphics[width=1.1\linewidth, page=4]{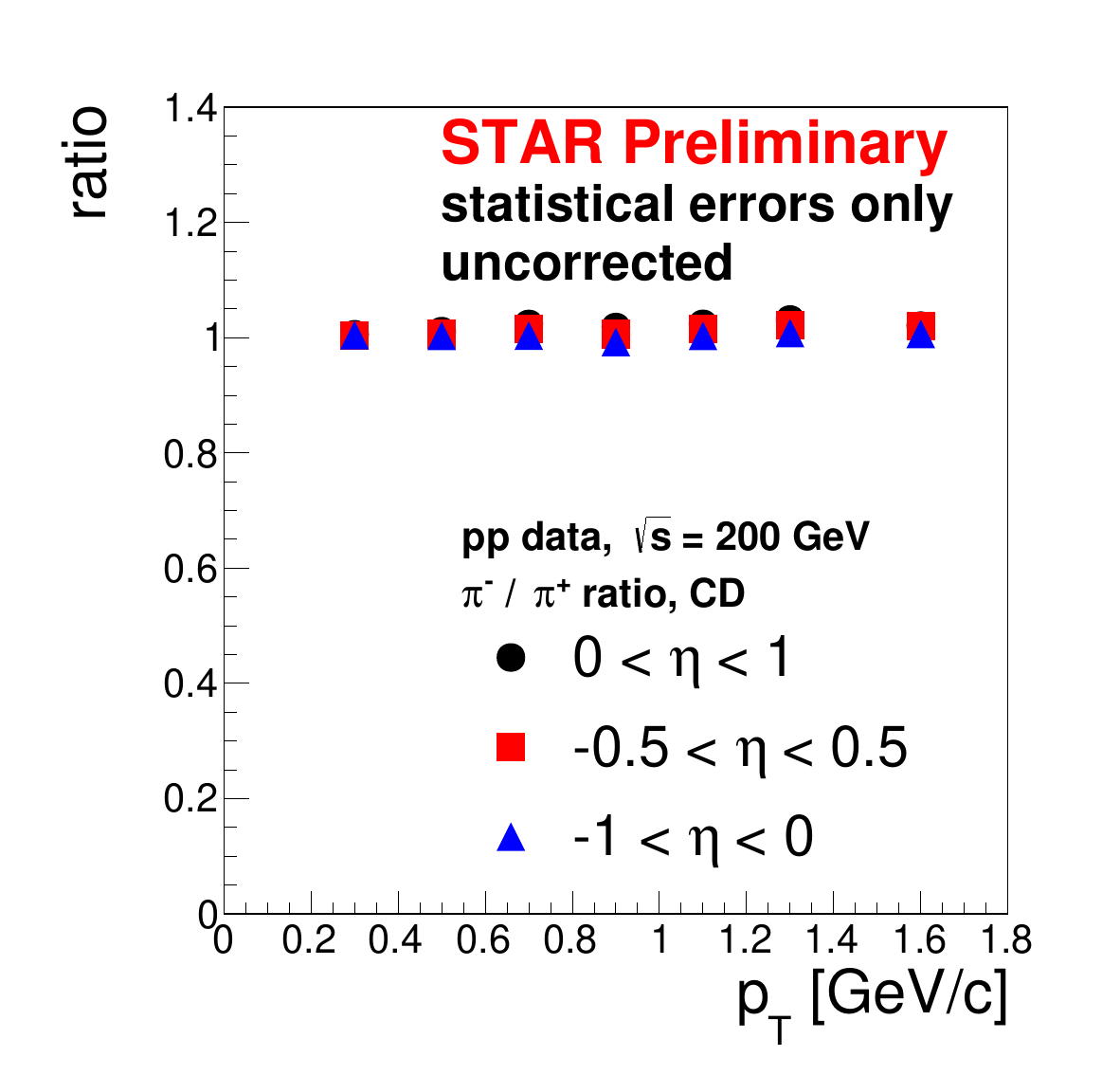}}}
		\end{subfigure}
	}
	\parbox{0.5\textwidth}{
		\centering
		\begin{subfigure}[b]{\linewidth}{
				{\includegraphics[width=1.1\linewidth, page=6]{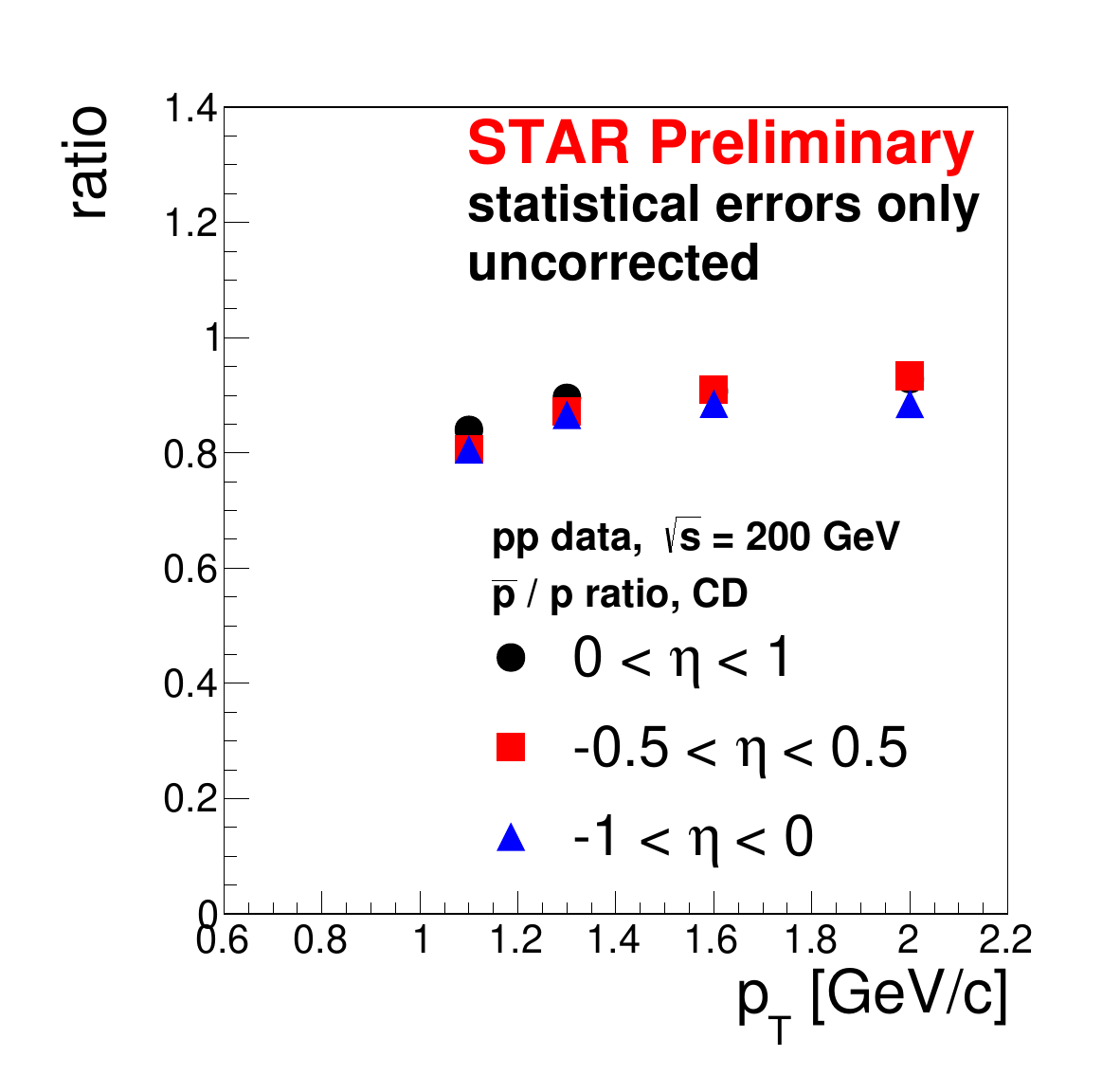}}}
		\end{subfigure}
	}
	
	\caption{The $\pi^-/\pi^+$ (left) and $\bar{p}/p$ (right) ratios measured at three pseudorapidity ranges in CD process. }
	\label{fig:ratiosCD}
\end{figure}

\begin{figure}
	\hspace*{-1em}
	\parbox{0.5\textwidth}{
		\centering
		\begin{subfigure}[b]{\linewidth}{
				{\includegraphics[width=1.1\linewidth, page=1]{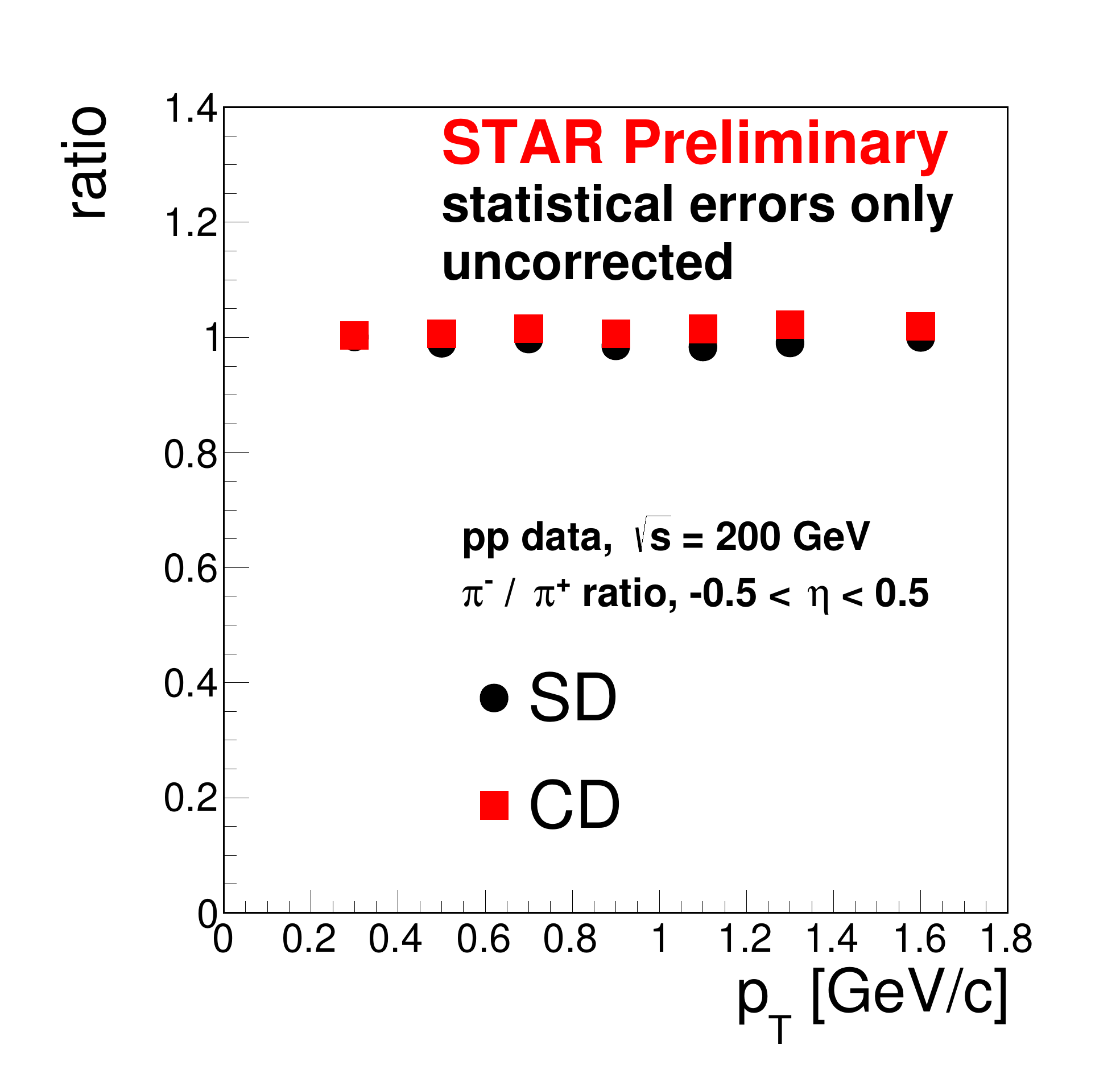}}}
		\end{subfigure}
	}
	\parbox{0.5\textwidth}{
		\centering
		\begin{subfigure}[b]{\linewidth}{
				{\includegraphics[width=1.1\linewidth, page=3]{pics/compareSDCDcentral.pdf}}}
		\end{subfigure}
	}
	
	\caption{Comparison of the $\pi^-/\pi^+$ (left) and $\bar{p}/p$ (right) ratios in $|\eta| < 0.5$ interval between CD and SD processes.}
	\label{fig:ratiosCDandSD}
\end{figure}

\begin{figure}
	\hspace*{-1em}
	\parbox{0.5\textwidth}{
		\centering
		\begin{subfigure}[b]{\linewidth}{
				{\includegraphics[width=1.1\linewidth, page=1]{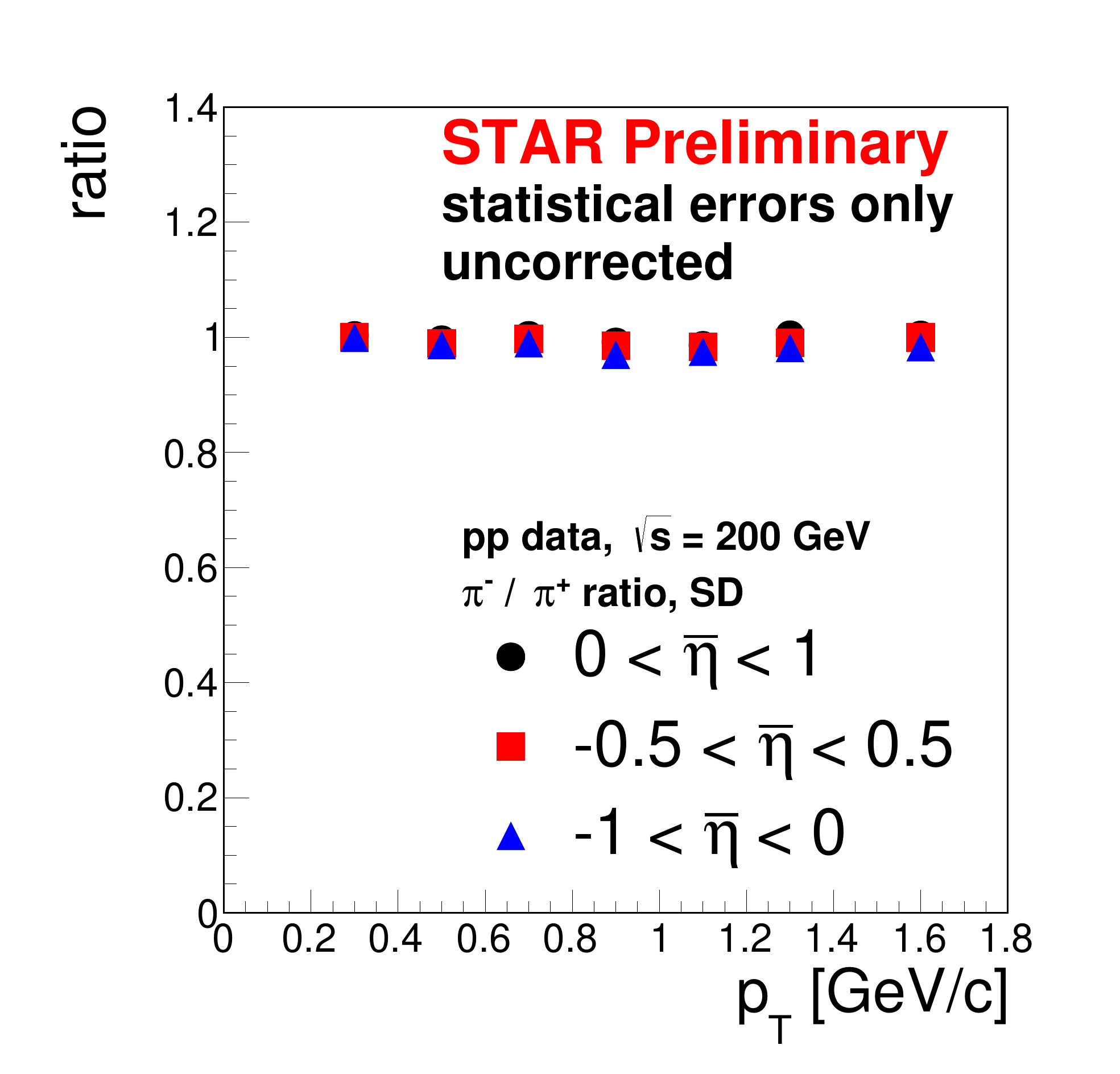}}}
		\end{subfigure}
	}
	\parbox{0.5\textwidth}{
		\centering
		\begin{subfigure}[b]{\linewidth}{
				{\includegraphics[width=1.1\linewidth, page=3]{pics/compareEtaCorr.pdf}}}
		\end{subfigure}
	}
	\caption{The particles ratio measured in SD events for three $\bar{\eta}$ bins. The $\bar{p}/p$ ratio in SD is corrected for the $\bar{p}$ absorption (right) using CD.}
	\label{fig:ratiosSD}
\end{figure}

The particle to antiparticle ratios for pions and protons are shown in Fig.~\ref{fig:ratiosCD}  for CD process and in Fig.~\ref{fig:ratiosSD} for SD process. The comparison of particle-antiparticle ratios between SD and CD in most central pseudorapidity interval $-0.5 < \eta < 0.5$ are presented in Fig.~\ref{fig:ratiosCDandSD}.  Charged pions are identified with transverse momentum $0.2 \leq p_T \leq 1.8$ GeV/c. In both processes, the $\pi^-/\pi^+$  ratio is equal to 1 in almost all $p_T$ bins and all $\eta\left(\bar{\eta}\right)$ ranges . This result is also consistent with the STAR minimum bias inelastic measurements~\cite{PhysRevC.79.034909}. The $\bar{p}/p$ ratio is analyzed for particles with $1 < p_T < 2.2$~GeV/c because in that $p_T$ range the background from secondary interactions is expected to be small.   Due to the antiproton absorption, the TPC track reconstruction efficiency is smaller for  antiprotons than that for protons. Hence, the $\bar{p}/p$ ratio is below 1 in CD, but the same in all three $\eta$ ranges. Nevertheless, it is noticed that this ratio is smaller in SD than that in CD. The $\bar{p}/p$ ratio in SD divided by the $\bar{p}/p$ ratio in CD, $\left[\bar{p}/p(\textnormal{SD})\right]/\left[\bar{p}/p(\textnormal{CD})\right]$, gives the ratio in SD corrected for $p$, $\bar{p}$ efficiencies (Fig.~\ref{fig:ratiosSD}). As a result, the $\bar{p}/p$ ratio in the SD for particles within $-0.5 < \bar{\eta} < 0.5 $ varies between $0.9$ and $0.95$. And the ratio is greater than that in STAR minimum bias inelastic measurements, where it is between $0.75$ and $0.85$. Moreover, the comparison of the $\bar{p}/p$ ratio in different $\bar{\eta}$ ranges indicates that baryon number transfer is smaller in the outgoing proton direction.

\section{Summary}
The distributions of charged-particle multiplicity, $p_T$ and $\eta$ measured in SD and CD processes are not well reproduced by PYTHIA~8 simulations in full accessible ranges.
The preliminary results on $\pi^-/\pi^+$ ratios are in agreement with the STAR MB inelastic measurements, while the~$\bar{p}/p$ ratio in SD is measured to be greater than that in MB inelastic $p$+$p$ collisions.


\begin{thebibliography}{15}

\bibitem{cms}
V. Khachatryan \textit{et al.} (CMS Collab.), Phys. Lett. B 751, 143 (2015).
\bibitem{atlaschrg}
G. Aad \textit{et al.} (ATLAS Collab.), New J. Phys. 13, 053033 (2011).
\bibitem{PhysRevC.79.034909}
B. I. Abelev \textit{et al.} (STAR Collab.), Phys. Rev. C 79, 034909 (2009).
\bibitem{Adam:1999164}
J. Adam \textit{et al.} (ALICE Collab.), Eur. Phys. J. C 75, 226 (2015).
\bibitem{Kopeliovich:1988qm}
B. Z. Kopeliovich, B. G. Zakharov, Z. Phys. C 43, 241 (1989).
\bibitem{Rossi:1977cy}
G. C. Rossi, G. Veneziano, Nucl. Phys. B 123, 507-545 (1977).
\bibitem{Aamodt:2010dx}
K. Aamodt \textit{et al.} (ALICE Collab.), Phys. Rev. Lett. 105, 072002 (2010).
\bibitem{Ackermann:2002ad}
K.  Ackermann \textit{et al.} (STAR Collab.), Nucl. Instrum. Meth. A 499, 624 (2003).
\bibitem{wlodek}
W. Guryn, Proceedings of this conference (Diffraction and Low-x 2018).
\bibitem{pythia8}
T. Sjostrand, S. Mrenna, P. Skands, Comput. Phys. Commun. 178, 852 (2008).
\bibitem{Schuler:1993wr}
G. A. Schuler, T. Sjostrand, Phys. Rev. D 49, 2257 (1994).
\bibitem{Ciesielski}
R. Ciesielski, K. Goulianos, PoS ICHEP2012, 301 (2013).
\end{thebibliography}
\end{document}